\documentclass[journal,10pt]{IEEEtran}

\usepackage{newtxtext,newtxmath}
\usepackage{cite}
\usepackage{amsmath,amsfonts}
\usepackage{algorithmic}
\usepackage{graphicx}
\graphicspath{{figures/}} 
\usepackage{textcomp}
\usepackage{xcolor}
\usepackage{nomencl}
\makenomenclature
\usepackage{physics}

\usepackage{anyfontsize}
\usepackage[]{ragged2e}
\usepackage{tablefootnote}
\usepackage{hyperref}
\usepackage{makecell}
\usepackage{threeparttable}

\usepackage{array}
\newcolumntype{C}{>{\centering\arraybackslash}m{27mm}}
\newcolumntype{B}{>{\centering\arraybackslash}m{12mm}}
\newcolumntype{A}{>{\centering\arraybackslash}m{8mm}}

\begin{document}

\title{Performance of Particle Tracking Using\\ a Quantum Graph Neural Network}

\DeclareRobustCommand*{\IEEEauthorrefmarkN}[1]{%
  \raisebox{0pt}[0pt][0pt]{\textsuperscript{\footnotesize #1}}%
}

\author{
Cenk~T\"uys\"uz\IEEEauthorrefmarkN{1}\IEEEauthorrefmark{2}\IEEEauthorrefmark{3},
Kristiane~Novotny\IEEEauthorrefmarkN{4},
Carla~Rieger\IEEEauthorrefmarkN{2},
Federico Carminati\IEEEauthorrefmarkN{5},
Bilge Demirk\"oz\IEEEauthorrefmarkN{1}\\
Daniel Dobos\IEEEauthorrefmarkN{4}\IEEEauthorrefmark{5},
Fabio Fracas\IEEEauthorrefmarkN{5}\IEEEauthorrefmark{6},
Karolos Potamianos\IEEEauthorrefmarkN{4}\IEEEauthorrefmark{1},
Sofia Vallecorsa\IEEEauthorrefmarkN{5}
~and~Jean-Roch~Vlimant\IEEEauthorrefmarkN{3}\\[2ex]

\IEEEauthorrefmarkN{1}{\textit{Middle East Technical University, Ankara, Turkey}\\}
\IEEEauthorrefmarkN{2}{\textit{ETH Zurich, Zurich, Switzerland}\\}
\IEEEauthorrefmarkN{3}{\textit{California Institute of Technology, Pasadena, California, USA}\\}
\IEEEauthorrefmarkN{4}{\textit{gluoNNet, Geneva, Switzerland}\\}
\IEEEauthorrefmarkN{5}{\textit{CERN, Geneva, Switzerland \vspace{-18pt}}}

\thanks{\IEEEauthorrefmark{2}cenk.tuysuz@cern.ch}
\thanks{\IEEEauthorrefmark{3}Also with STB Research,\IEEEauthorrefmark{5}Also with Lancaster University,\IEEEauthorrefmark{1}Also with University of Oxford,\IEEEauthorrefmark{6}Also with University of Padua.}
}

\markboth{Başarım 2020 - High Performance Computing Conference}%
{}

\maketitle

\begin{abstract}
The Large Hadron Collider (LHC) at the European Organisation for Nuclear Research (CERN) will be upgraded to further increase the instantaneous rate of particle collisions (luminosity) and become the High Luminosity LHC. This increase in luminosity, will yield many more detector hits (occupancy), and thus measurements will pose a challenge to track reconstruction algorithms being responsible to determine particle trajectories from those hits. This work explores the possibility of converting a novel Graph Neural Network model, that proven itself for the track reconstruction task, to a Hybrid Graph Neural Network in order to benefit the exponentially growing Hilbert Space. Several Parametrized Quantum Circuits (PQC) are tested and their performance against the classical approach is compared. We show that the hybrid model can perform similar to the classical approach. We also present a future road map to further increase the performance of the current hybrid model.
\end{abstract}


%
\IEEEpeerreviewmaketitle

\nomenclature{$p_{T}$}{Transverse (xy plane) momentum}
\nomenclature{$\phi$}{Angle in transverse plane}
\nomenclature{$z$}{Particle translation axis before collision}
\nomenclature{$\eta$}{Pseudorapidity, a measure of angle w.r.t. z axis}
\printnomenclature

\section{Introduction}
\noindent Particle collider experiments aim to understand the nature of particles by colliding groups of particles at high energies and try to observe the processes to validate the theories. The Large Hadron Collider (LHC) at CERN serves to several experiments where each specializes to observe certain processes. These experiments require to be highly sensitive and therefore, need sophisticated software and hardware.\\
\indent Additional to requirements above, these experiments need very fast processing units as the time between two bunch collisions is very short (reaching up to 1 MHz for ATLAS~\cite{atlas-report} and CMS~\cite{cms-report}). A big data storage and processing problem arises, when the fast data acquisition is combined with sensitive hardware~\cite{lhc-computing}. A total disk and tape space of 990 PB and around 550k CPU cores were pledged to LHC experiments in 2017 according to a report by CERN Computing Resources Scrutiny Group (CRSG)~\cite{computing-report}.
\indent Currently, LHC is going through an upgrade period to increase the luminosity of the particle beams~\cite{ref-hilumi}. Therefore, the future High Luminosity Large Hadron Collider (HL-LHC) experiments will require much faster electronics and software to process the increased rate of collisions. \\
\indent Particle track reconstruction problem is one of the challenges that becomes important with the HL-LHC upgrade. In this problem, the aim is to identify the trajectory of particles using the measurements of the tracking detectors. Although, there are novel algorithms that can handle the current rate of collisions, they suffer from high collision rates as they scale worse than polynomially. \\
\indent Recent developments in Quantum Computing (QC) allowed scientists to look at computational problems from a new perspective. There is a great effort to make use of these new tools provided by QC to gain speed-ups for many computational tasks in High Energy Physics as well~\cite{qml-hep}. Since the particle track reconstruction problem also suffers from scaling, researchers are looking at QC tools to improve the scaling. While there are several attempts using adiabatic QC~\cite{qalg1,qalg2,qalg3}, this work focuses on gate-type Quantum Computing. \\
\indent In this work, we try to improve on our initial work, where we investigated the use of a Quantum Graph Neural Network approach to solve the particle track reconstruction problem~\cite{tuysuz-1,tuysuz-2}. We present an analysis of several well-performing Quantum Circuits and give a comparison with the novel classical equivalent, which our approach is based on. \\
\indent The rest of the paper is as follows. Details of the dataset used and pre-processing methods employed are given in Chapter~\ref{chapter:dataset}. 
The Hybrid Graph Neural Network Model employed is explained in detail in Chapter~\ref{chapter:qgnn}. Results and comparison with novel methods are given in Chapter~\ref{chapter:results}. Finally, a discussion of results and comments on possible improvements are given in Chapter~\ref{chapter:conclusion}.

\section{The Dataset and Pre-processing} \label{chapter:dataset}
\noindent Particles are accelerated and then collide in bunches at particle colliders. These bunch collisions happen at every 25 ns where approximately 20 protons collide at the
current LHC. These collisions create new particles that scatter through all directions. A cylinder-shaped tracking detector is usually at

\noindent the core of these types of experiments. The aim of these detectors is to record signals that are called \textit{hits}, which are created when a particle passes. Then, an algorithm is used to distinguish hits that belong to each particle. This task is referred to as particle track reconstruction.

CERN organized the Kaggle Tracking Machine Learning (TrackML) challenge to invite Machine Learning (ML) experts to solve the problems in particle track reconstruction in 2018~\cite{trackml}. Since then, the dataset is being used for many researchers to benchmark their results. This work follows the trend and uses the TrackML dataset to give a better comparison to other approaches.

\begin{figure}[htbp]
\centerline{\includegraphics[width=\linewidth]{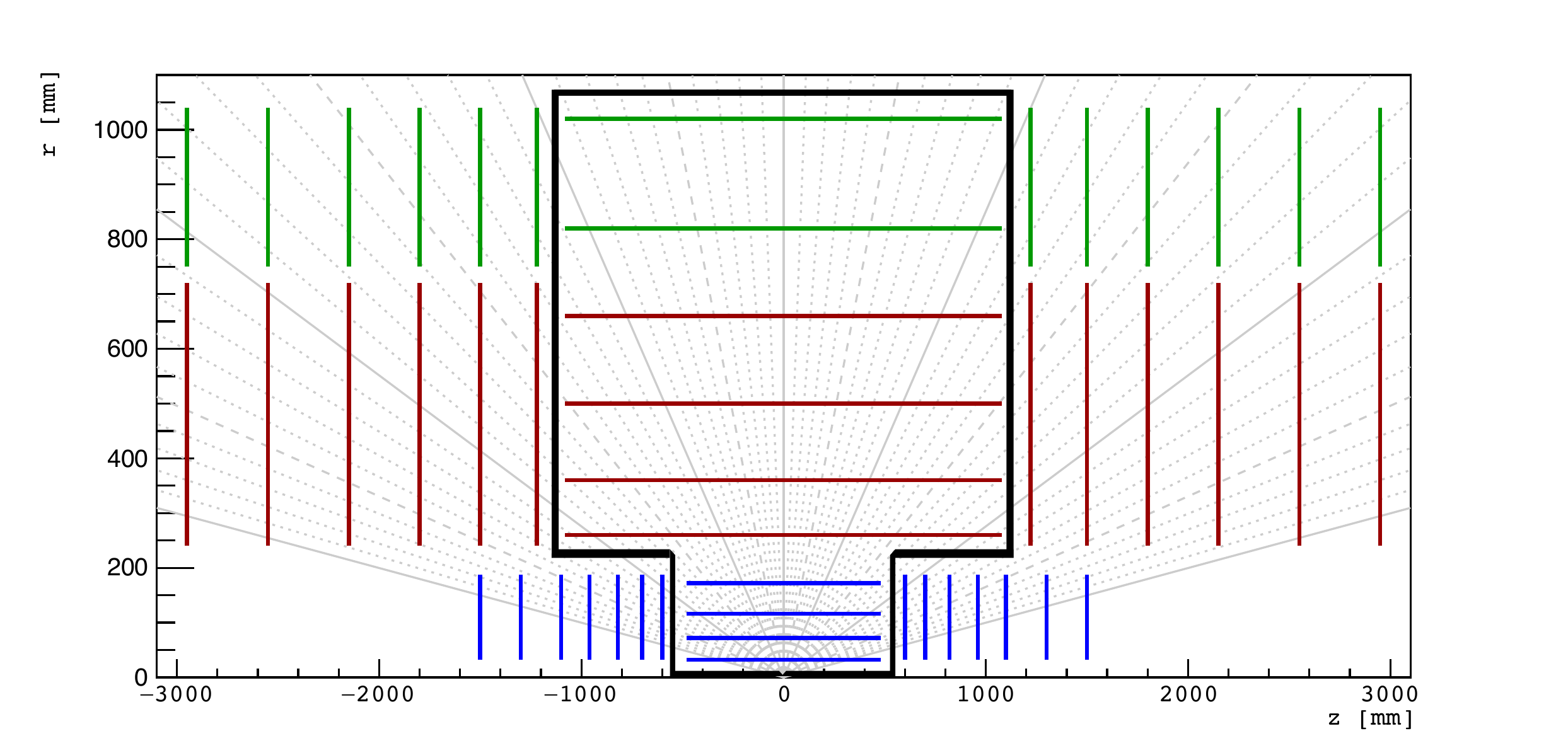}}
\caption{TrackML detector geometry and the selected barrel region.}
\label{fig:detector}
\end{figure}

The TrackML dataset contains 10k simulated events. In this work, only 100 events representing 1\% of the available dataset are used due to constraints in run time of quantum circuit simulations which may take several weeks. To further reduce the size of the dataset, we omit some parts of the detector to decrease the high combinatorics and track ambiguity caused by vertical (endcap) layers. Horizontal (barrel) layers (8, 13, 17) that are located near the center of the detector are kept, while the data coming from vertical (endcap) layers are omitted. The geometry of the TrackML detector that is used to create the simulated dataset, as well the part of the detector that is selected to be used in this work can be seen in Fig.~\ref{fig:detector}. 

The TrackML dataset contains spatial coordinate information as well as the particle id of each hit. The dataset is used to construct a preliminary graph that connects all the relevant hits, while also preserving the ground truth information by checking the particle id's that belong to same particles. These connections are made using a set of selections that can be seen in Table.~\ref{table:cuts} to keep pre-processing time short and to keep the selection process as abstract as possible. 

These selection choices are the same as used by HepTrkX~\cite{HepTrkX} and their  pre-processing code is used, as well. Since particles move along the $+\hat{r}$ direction we can slice the graphs without loosing much information. All events are sliced to 8 pieces in $\eta$ and to 2 pieces in $z$ directions to reduce the size of each graph. Although there are tracks lost near the edges of the slices, they are omitted in this work. Also there are proposed ideas to recover those lost tracks by coinciding the slices, however this procedure is not a topic of this work.

\begin{figure}[htbp]
\centerline{
\includegraphics[width=0.7\linewidth]{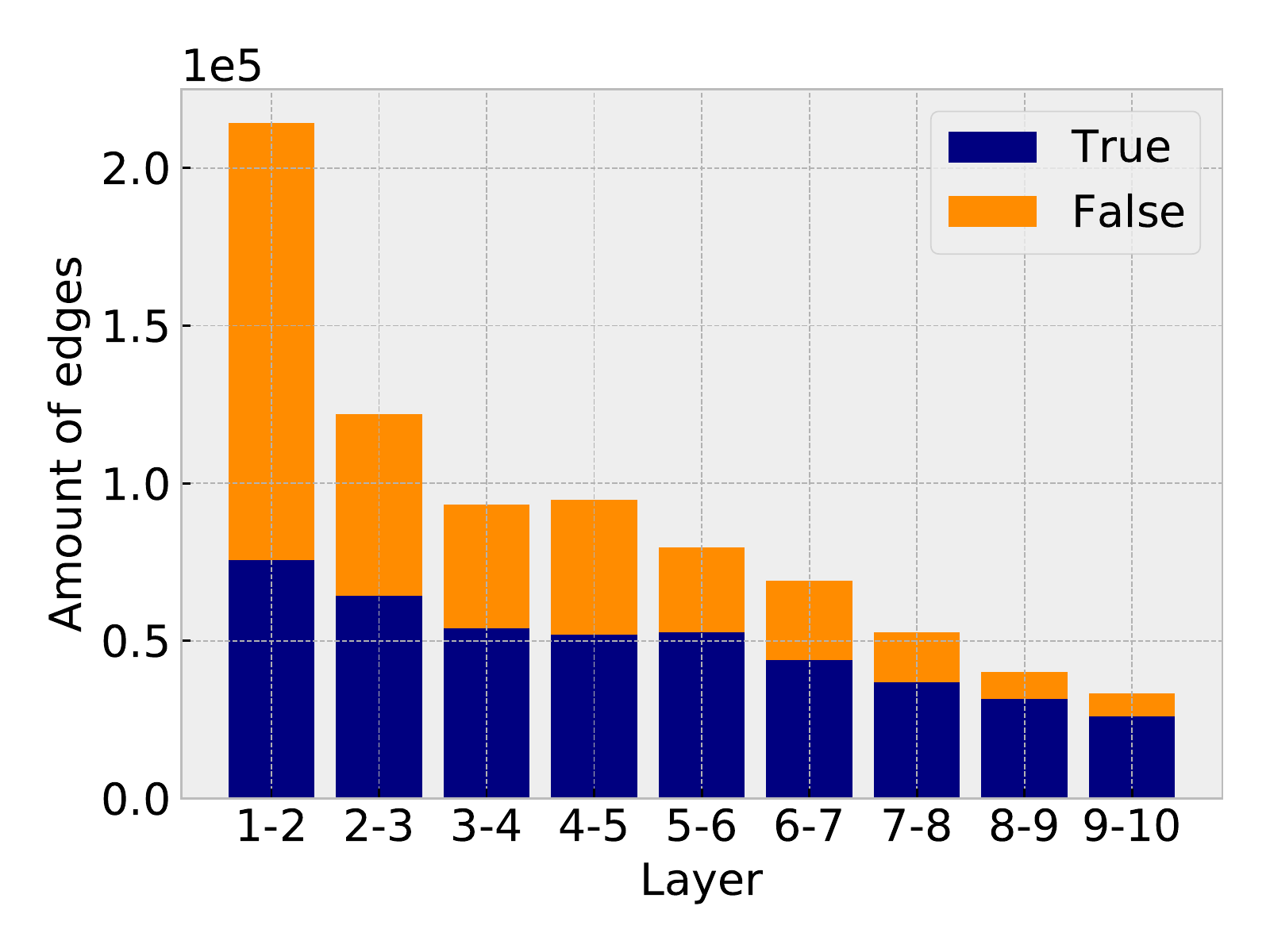}
}
\caption{Bar plot of edges by layers of a single event that corresponds to 16 subgprahs. The change in ratio of True (blue) / Fake (orange) edges per layer shows that combinatorics creates many fake edges, particularly in the initial layers.}
\label{fig:edge-dist}
\end{figure}

\begin{figure}[!h]
\centerline{\includegraphics[width=\linewidth]{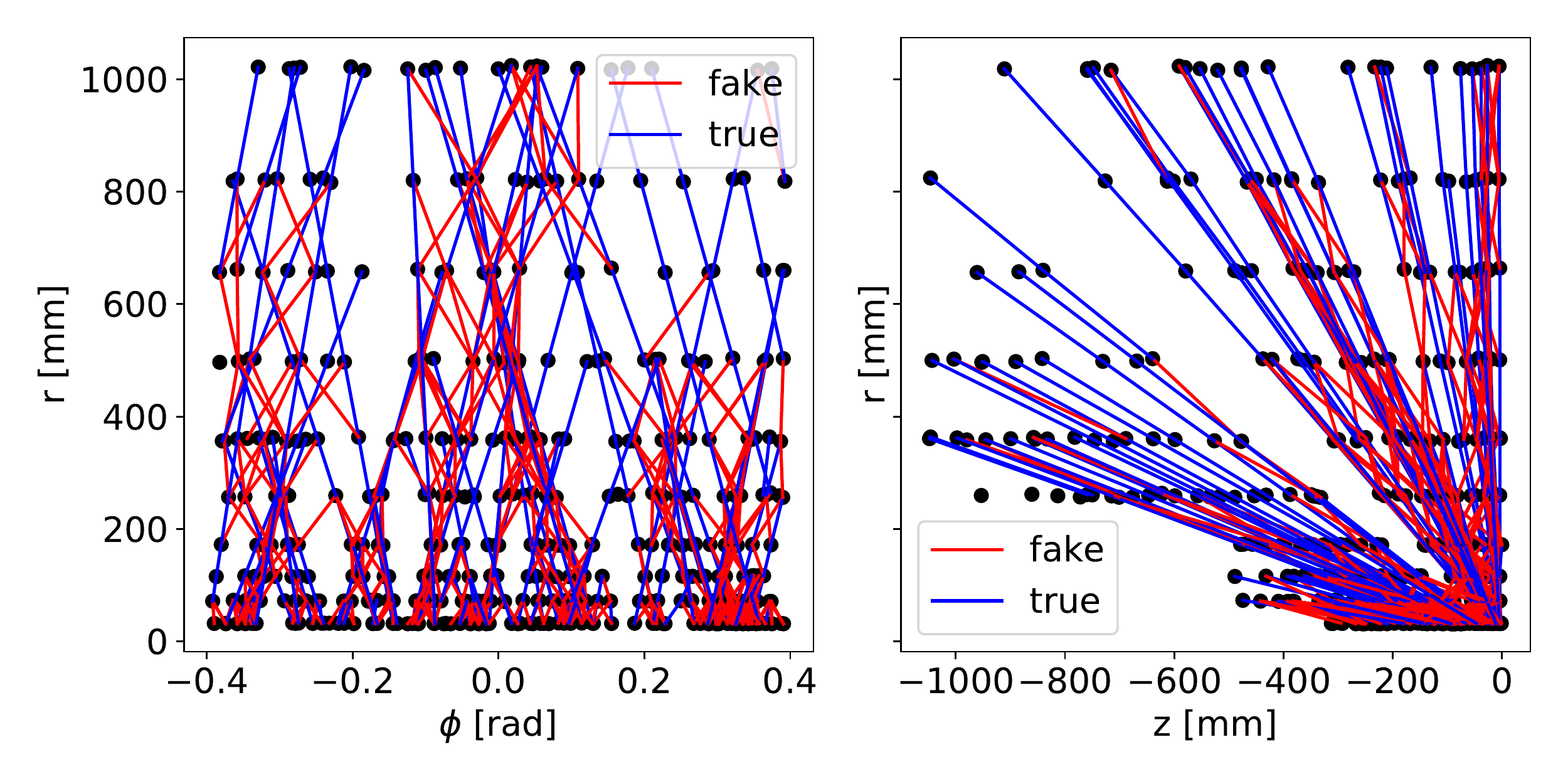}}
\caption{One subgraph (1/16 of an event) after preprocessing in cylindrical coordinates. On the left plot, r vs $\phi$ is shown. On the right, the same subgraph is displayed in r vs. z coordinates. True edges are displayed in blue, while false edges are shown in red.}
\label{fig:cylindrical}
\end{figure}

After pre-processing, the dataset contains 1600 subgraphs which were constructed using 100 events, where hits are now called \textit{nodes} and tracks that connect hits are called as \textit{edges}. A subgraph contains 350 nodes and 500 edges on average. The change of edges by layers can be seen in  Fig.~\ref{fig:edge-dist} and a subgraph can be seen in Fig.\ref{fig:cylindrical}. 

\begin{table}[htbp]
  \begin{center}
      \caption{Selections applied to TrackML dataset for preprocessing.}
    \begin{tabular}{|c|c|}
\hline
Parameter & Constraint     \\\hline
$\left|p_T\right|$ & $>1 GeV$     \\\hline
$\Delta\phi$ & $<0.0006$  \\\hline
$z_0$ & $<100 mm$  \\\hline
$\eta$ & $[-5,5]$  \\\hline
    \end{tabular}
    \label{table:cuts}
  \end{center}
\end{table}

\section{Quantum Graph Neural Network Model} \label{chapter:qgnn}

\begin{figure*}[htbp]
\centerline{\includegraphics[width=\textwidth]{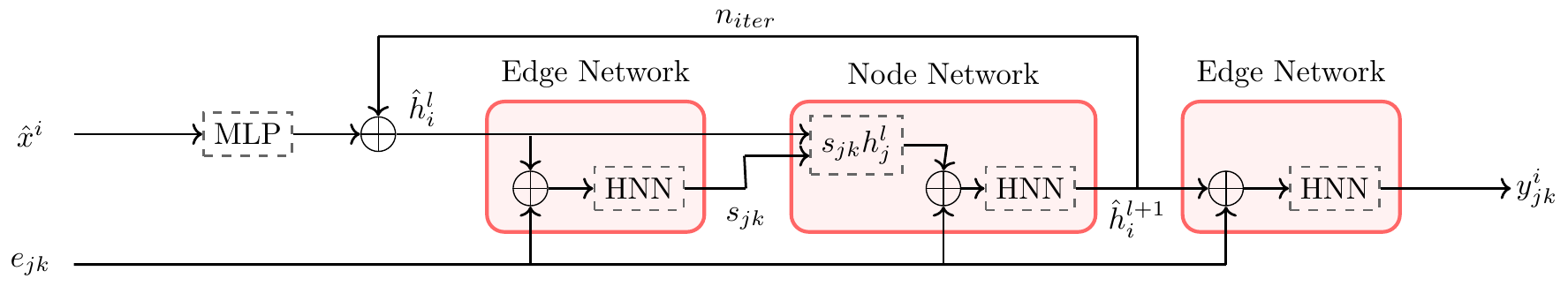}}
\caption{The overall Quantum Graph Neural Network pipeline.}
\label{fig:heptrkx-quantum}
\end{figure*}

\noindent The Quantum Graph Neural Network (QGNN) model takes a pre-processed subgraph as an input and gives a probability output for all initial edges. The model is built over an Attention Passing Graph Neural Network following the same strategy with the HepTrkX project, which is the classical equivalent of this work. Different than the classical approach, in this work, we replace some of the Multi Layer Perceptrons (MLP) with Quantum Neural Networks (QNN). Therefore, we achieve to build a Hybrid Graph Neural Network model that was previously referred to as QGNN. As the MLPs are directly converted to QNNs, the QGNN name is selected to be used, although the model is in fact a hybrid model. The overall pipeline is displayed in Fig.~\ref{fig:heptrkx-quantum}.

\subsection{Classical Parts of the Network} 
\noindent The QGNN consists of 4 important building blocks. First is the input layer MLP, whose task is to increase the dimension of the input data to a higher dimension. Input layer takes the spatial coordinate information as input and concatenates the required hidden dimensions to the input. Therefore, the output of the input layer is the first node feature vector and it is referred to as $h_i^0$, where $h_i^l \in {\rm I\!R}^{3+N_{hidden}}$.

The second building block is the Edge Network. Edge Network takes the node features and uses the connectivity matrix ($e_{jk}$) to select all initially connected edges that are also called as doublets. Then, a QNN is run for all edges, whose output is the edge features ($s_{jk}$). Edge features are the probabilities of the edges considered as true, hence $s_{jk} \in {\rm I\!R}^{1}$.

The edge features are used by the third building block, which is the Node Network. Node Network is very similar to Edge Network. However, it takes the new edge features and uses them to update node features. The $s_{jk}$ matrix is used to construct the weights between all nodes. Then, a combination of all nodes with its neighbours are created. Therefore, this neighbour nodes array consists of combinations of 3 nodes that are called triplets. Finally, the triplets array is fed to the QNN, but this time the output would be the new node features ($h_i^{l+1}$). Then, the output is fed to the Edge Network again, and this recurrent process is repeated $\text{N}_{\text{iterations}}$ times.

\subsection{The Quantum Neural Network}
\noindent The fourth and most important building block is the QNN. As mentioned above, the QNN is used inside both the Edge Network and the Node Network. However, the dimension of the input is different for each case. 

QNN of the Edge Network takes $N_{doublets}$ inputs with size ${\rm I\!R}^{2\times(3+N_{hidden})}$. The QNN takes each edge as input and produce $s_{jk} \in {\rm I\!R}^{1}$. QNN of the Node Network takes $N_{triplets}$ inputs with size ${\rm I\!R}^{3\times(3+N_{hidden})}$. And, the output is $h_i^{l+1} \in {\rm I\!R}^{N_{hidden}}$. The discrepancy in the input size requires use of two different QNNs.

\begin{figure}[!ht]
\centerline{\includegraphics[width=\linewidth]{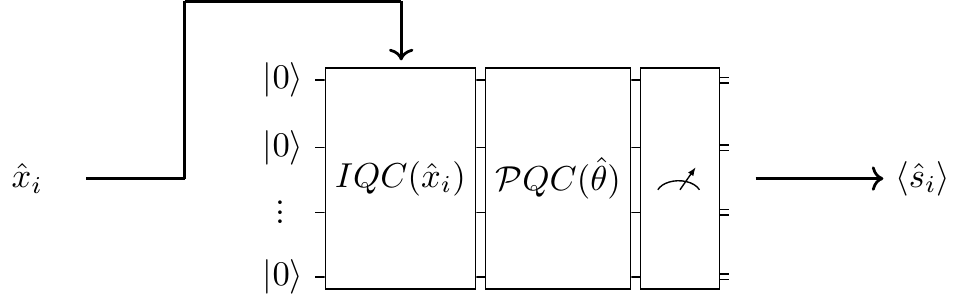}}
\caption{The QNN data flow diagram.}
\label{fig:QNN}
\end{figure}

The QNN has three layers that can be seen in Fig.~\ref{fig:QNN}. First layer, which is called Information Encoding Quantum Circuit (IQC), is used to encode the input data to qubits. The second layer is the PQC, which has its own trainable parameters. The final layer is the measurement layer where the states of the qubits are observed.  

\subsubsection{Information Encoding}
\begin{figure}[!htbp]
\centerline{\includegraphics[]{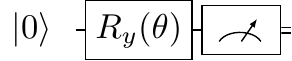}}
\caption{Information encoding using a single qubit.}
\label{fig:IQC1}
\end{figure}

\noindent The first layer of the QNN block is responsible for information encoding. A very careful approach should be followed to make use of the exponentially growing Hilbert Space when encoding the classical data. A smart choice of encoding can allow the model to avoid barren plateaus and speed-up the training \cite{barren,barren-2}. We follow a simple encoding strategy, as the width, depth and number of qubits in our model is relatively low. This strategy involves using only $N_{qubits}$ $R_y$ gates. A one qubit example is given below, where a single variable $\theta$ is encoded to a qubit. Then, the expectation value of the circuit w.r.t. to Z-direction is calculated to observe the encoded information vs. the classical information. These steps are given from \eqref{eq:encoding} to \eqref{eq:encoding3} and the quantum circuit representation is given in Fig.~\ref{fig:IQC1}.

\begin{align}
\ket{\Psi}_\theta &= R_Y(\theta) \ket{0} \nonumber\\
&= 
\begin{pmatrix}
\cos(\theta/2) & -\sin(\theta/2) \\
\sin(\theta/2) & \cos(\theta/2) 
\end{pmatrix} 
\begin{pmatrix}
1 \\
0
\end{pmatrix}\label{eq:encoding},\\ 
\ket{\Psi}_\theta &= \cos(\theta/2)\ket{0}+\sin(\theta/2)\ket{1},\\
\text{M} &= \bra{\Psi} \sigma_Z \ket{\Psi}_\theta\label{eq:encoding2},\\
\text{M} &= \cos(\theta/2)^2 - \sin(\theta/2)^2 \nonumber \\
 &= \cos\theta\label{eq:encoding3}
\end{align}

\begin{figure}[htbp]
\centerline{\includegraphics[]{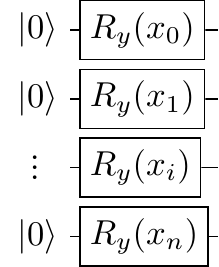}}
\caption{Information encoding quantum circuit.}
\label{fig:IQC}
\end{figure}

The $R_Y$ mapping produces a cos$\left(\theta\right)$ function, as it can be seen from~\eqref{eq:encoding3}. The cosine function is a periodic and not a 1-to-1 function. Therefore, only a half period of the function should be considered in order the avoid unwanted overlaps in data. Thus, we map the input data to be $\in$ [0,$\pi$]. This strategy is repeated for all data points. Therefore, for an input with the size ${\rm I\!R}^n$, there are n qubits needed. The extended information encoding quantum circuit (IQC) can be seen in Fig.~\ref{fig:IQC}.

\begin{figure*}[!ht]
\centering
\includegraphics[width=0.99\textwidth]{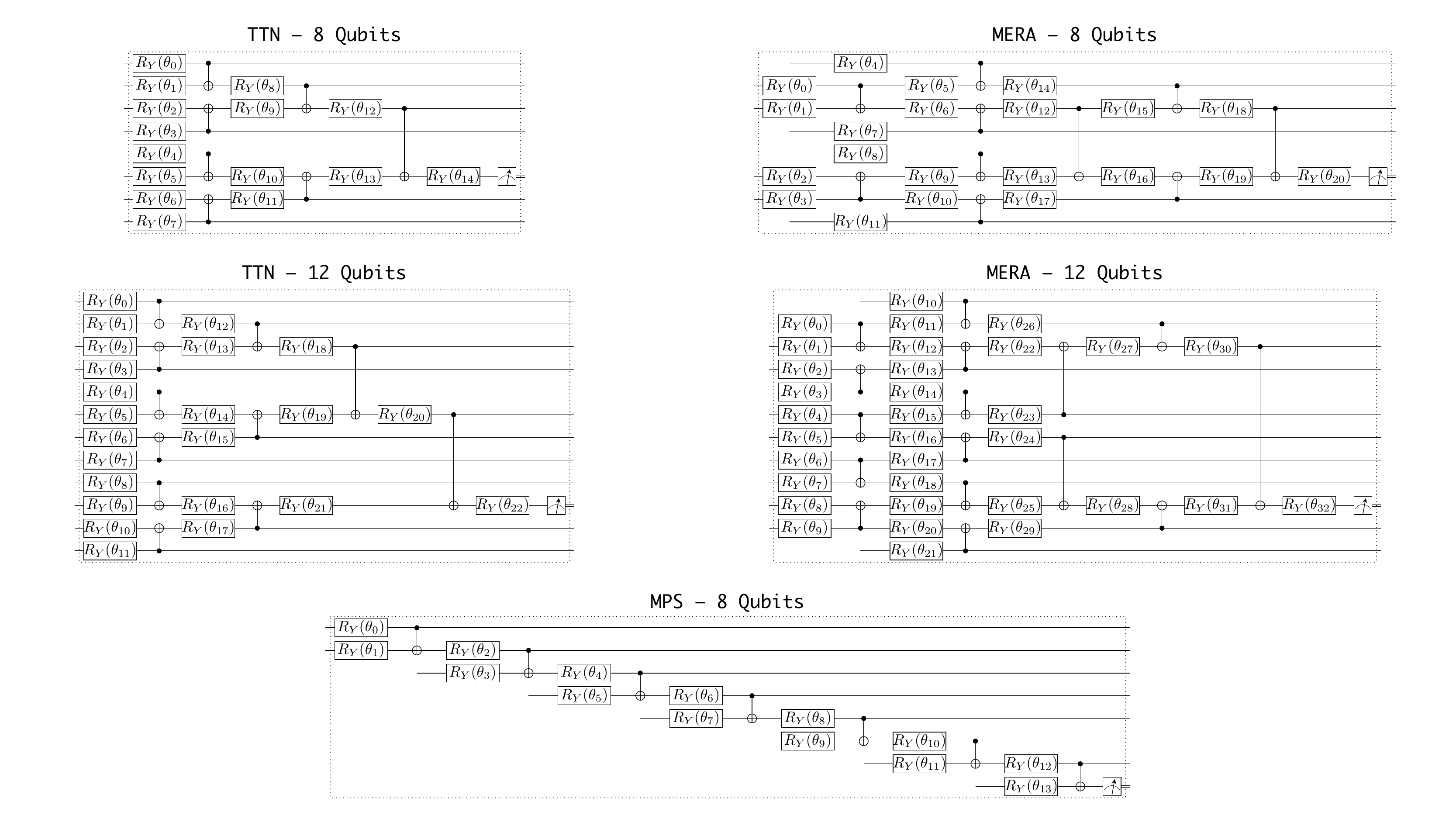}
\caption{Parametrized Quantum Circuits. Circuits are plotted using $\bra{q}\ket{pic}$~\cite{qpic}. They are based on the work by Bhatia~\cite{mps} and Grant~\cite{ttn}.}
\label{fig:PQC}
\end{figure*}

\subsubsection{Parametrized Quantum Circuits}
\noindent Parametrized Quantum Circuits (PQC) are quantum circuits whose operations depend on a list of parameters. PQCs lie at the heart of all Variational Quantum Algorithms. Choice of the PQC is very important as it determines the training performance. In this work, we use 3 PQC models, Matrix Product State (MPS)~\cite{mps}, tree tensor network (TTN) and multi-scale entanglement renormalization ansatz (MERA)~\cite{ttn}. These models are shallow and low width quantum circuits, which can be executed using Noisy Intermediate-Scale Quantum (NISQ) devices. Two different circuits are created for each model, as different number of qubits are needed for the Edge and Node Networks. These circuits use 8 qubits for the Edge Network and 12 qubits for the Node Network, when the hidden dimension size is 1. They can be seen in Fig.~\ref{fig:PQC}. Additionally, the parameters of the PQCs are always initialized with a uniform random function such that they are $\in$ [0,4$\pi$].

\subsubsection{ Measurements}
\noindent The measurement block of the QNN is the last layer, where the quantum information is converted back to classical. When a measurement is taken, we loose a considerable amount of information as the quantum states collapses to one of the eigenstates of the observable we are measuring. To avoid the loss of information, several quantum state tomography methods are suggested in literature. In this work, we follow a simple strategy where we run each circuit 1000 times. The results are averaged to obtain an expectation value. This operation allows us to get a value of M for each qubit as also shown in \eqref{eq:encoding3}. The list of expectation values are referred to as the output of the QNN block.

\section{Training the Network}
\noindent The overall QGNN model is a hybrid structure, meaning that it contains parts that needs to be executed on a regular computer as well as parts that requires quantum computers. We can use conventional algorithms (e.g. gradient descent) when training the classical parts of the model. However, there is a need for some additional steps when training a quantum model. Algorithms that calculates gradients over quantum circuits (e.g. parameter shift rules~\cite{parameter-shift}) generally requires re-evaluation of the same circuit with different parameters, making the gradient taking a costly operation. Given that access to quantum computing hardware is rather limited at the moment, only simulations of quantum circuits are used in this work. 

There are additional difficulties to appear when a training a quantum circuit model. Most important of them is the barren plateau problem, which prevents taking gradient signals from quantum circuits as their  depth and width grow. Therefore, quantum circuits require additional strategies in the training step. However, as the PQCs used in this work are shallow circuits, an extra step is not employed further. This leaves an open door for further optimization of the current model.

Tensorflow~\cite{ref-tensorflow} is used to create the neural network pipeline. Pennylane~\cite{pennylane}is used to construct the connections between quantum circuits and classical neural networks. Qulacs~\cite{qulacs} is used to perform the quantum circuit simulations as it is one of the best performing quantum circuit simulator software. Qulacs allows use of GPUs as well as multi-thread CPU simulations which boosts the simulation times significantly. Experiments were performed using both GPU and multi-thread CPU options.

The dataset contains 1600 subgraphs, where 200 of them is selected randomly to be the validation set. All runs used the same validation set, while the order of the subgraphs is shuffled at each run, also all runs are repeated 3 times. Results in Chapter~\ref{chapter:results} include averages of these 3 runs. ADAM optimizer and binary cross entropy loss function of Tensorflow with a learning rate of 0.03 is used. The loss function is weighted according to amount of true and false edges, in order to avoid preference towards false edges. All models are trained only for 1 epoch due to constraints in run time. Total training time of models are given in Table~\ref{table:runtime}.

\begin{table}[htbp]
  \begin{center}
      \caption{Training times for different models.}
    \begin{tabular}{|B|B|C|B|}
\hline
\thead{PQC} &
\thead{Total\\ $\text{N}_{\text{parameters}}$} & 
\thead{Average Training Time\\ (1 training step)\\ $[minutes]$ } & \thead{CPU/GPU}\\[2ex]
\hline
\thead{MPS}  & \thead{40} & \thead{$2.43 \pm 0.76$} & \thead{CPU\footnotemark}\\
\hline
\thead{TTN}  & \thead{42} & \thead{$3.94 \pm 1.43$}& \thead{GPU\footnotemark}\\
\hline
\thead{MERA} & \thead{58}  & \thead{$6.45 \pm 2.38$} & \thead{GPU}\\
\hline
\end{tabular}
\begin{tablenotes}
         \smallskip
     \item $^1$Intel(R) Xeon(R) Silver 4216 CPU @ 2.10GHz (8 threads)
     \item $^2$CPU (1 thread) + Nvidia 1080 Ti GPU
\end{tablenotes}
    \label{table:runtime}
  \end{center}
\end{table}

\section{Results} \label{chapter:results}
\noindent In this section, the learning curves are presented for different models. We use Area Under the Curve (AUC) as the primary metric, which is a common choice in many Deep Learning applications. As it name suggest AUC is the area under the Receiver Operating Characteristic (ROC) curve and ROC is the curve obtained by plotting true positive rate (TPR) against the false positive rate (FPR). The main reason to use AUC is to better asses the capability of models to distinguish True Positive (TP) and True Negative (TN) outputs. TP and TN are chosen according to a threshold, but this the AUC metric looks at all possible thresholds and evaluate the model performance for all possible thresholds. Therefore, AUC gives a better assessment on a model's performance compared to good old accuracy and precision metrics. As AUC is the area of a normalized curve, it can have values from 0 to 1, where 1 is the perfect score and 0.5 represents ideal randomness.

\begin{figure}[htbp]
\includegraphics[width=0.5\linewidth]{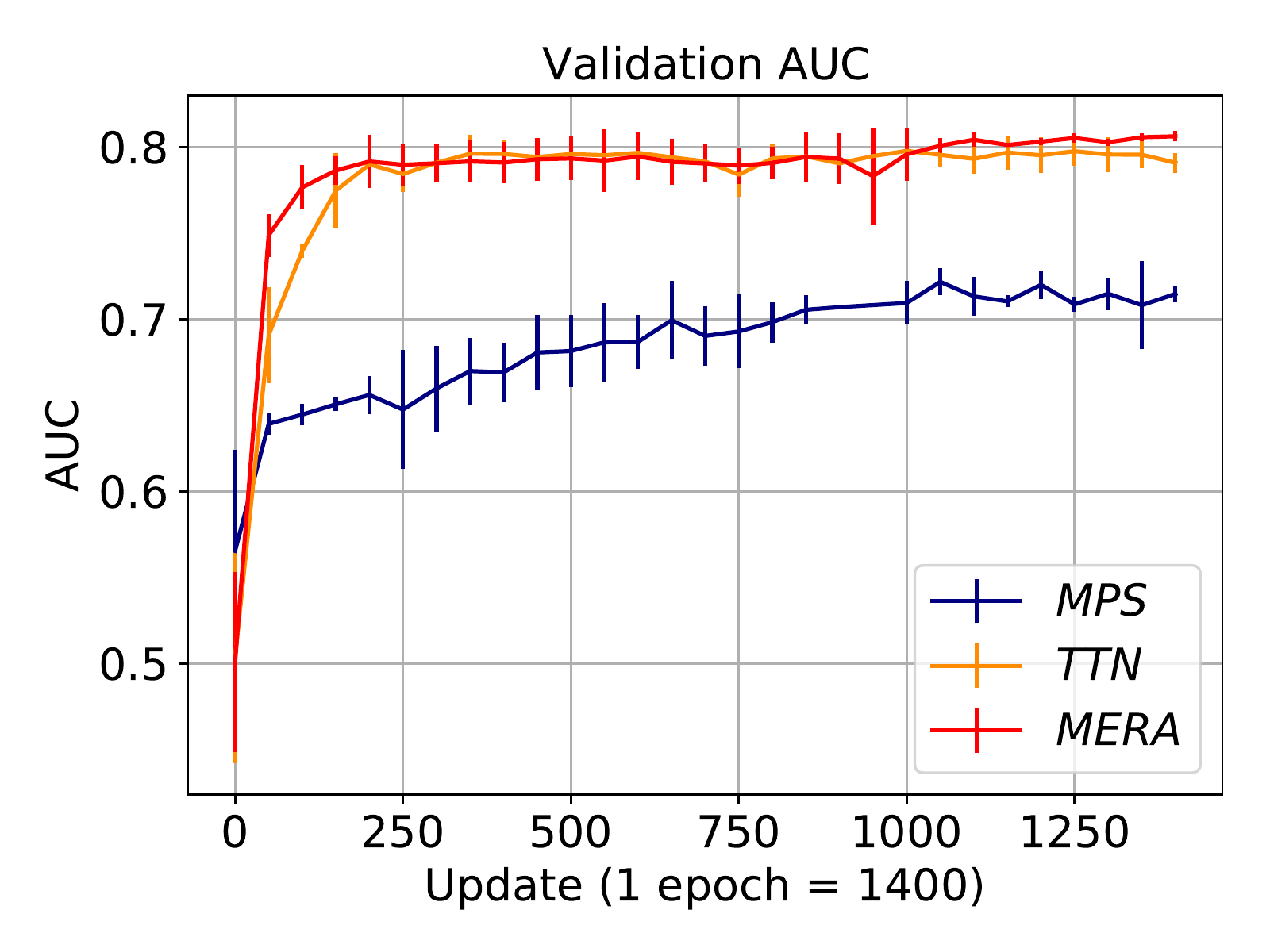}%
\includegraphics[width=0.5\linewidth]{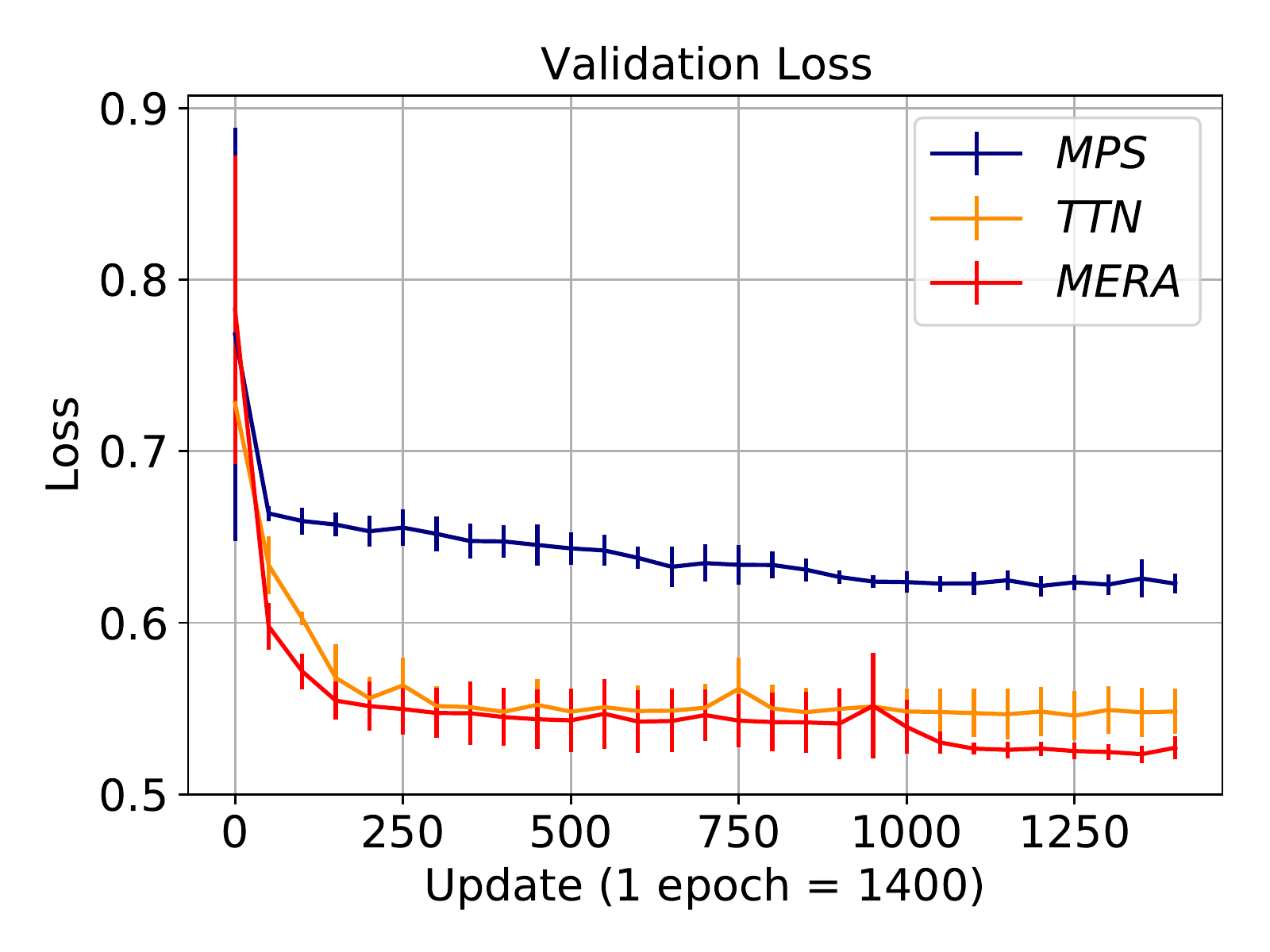}
\caption{Learning curves of the model with different PQCs. The AUC curve is plotted on the left. The loss curve is plotted on the right.}
\label{fig:results}
\end{figure}

We present the learning curves of the models for 1 epoch in Fig.~\ref{fig:results}. It can be seen from both plots that as the number of parameters of the model increase the performance gets better as expected. TTN and MERA models seem to perform close to 0.8 AUC. 

\begin{figure}[!ht]
\centerline{\includegraphics[width=0.8\linewidth]{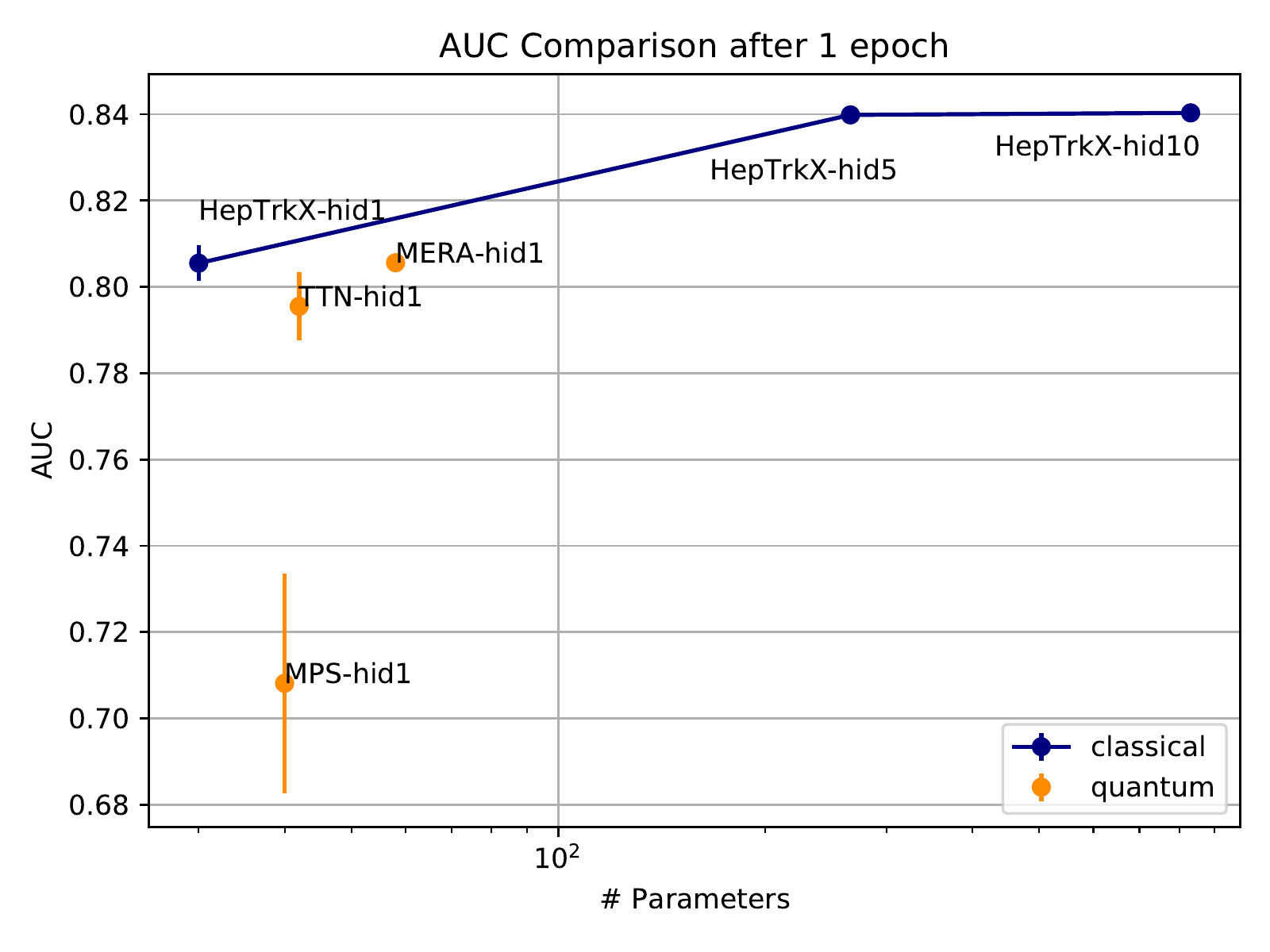}}
\caption{Comparison of results after 1 epoch against AUC and number of parameters.}
\label{fig:result-comp}
\end{figure}

The main training performance comes from iterating the model over the same training data, in Deep Learning. However, in our case, we only have 1 epoch. Therefore, we give a comparison with a simplified classical version after 1 epoch. The AUC values vs. number of parameters comparison after 1 epoch of training can be seen in Fig.~\ref{fig:result-comp}. Classical HepTrkX-GNN model is tested with hidden dimension sizes 1, 5 and 10, while our work only has 1 hidden dimension. All models are trained over 1 epoch using a single iteration.

We should note here that the original HepTrkX results suggest using a hidden dimension size of 128 with 6 iterations trained for 30 epochs. As classical models produce much larger gradients, we further decreased the learning rate from 0.03 to 0.001 only for the classical models. Thus, the model that we compare against here is a simplified version.

\begin{figure}[htbp]
\includegraphics[width=0.5\linewidth]{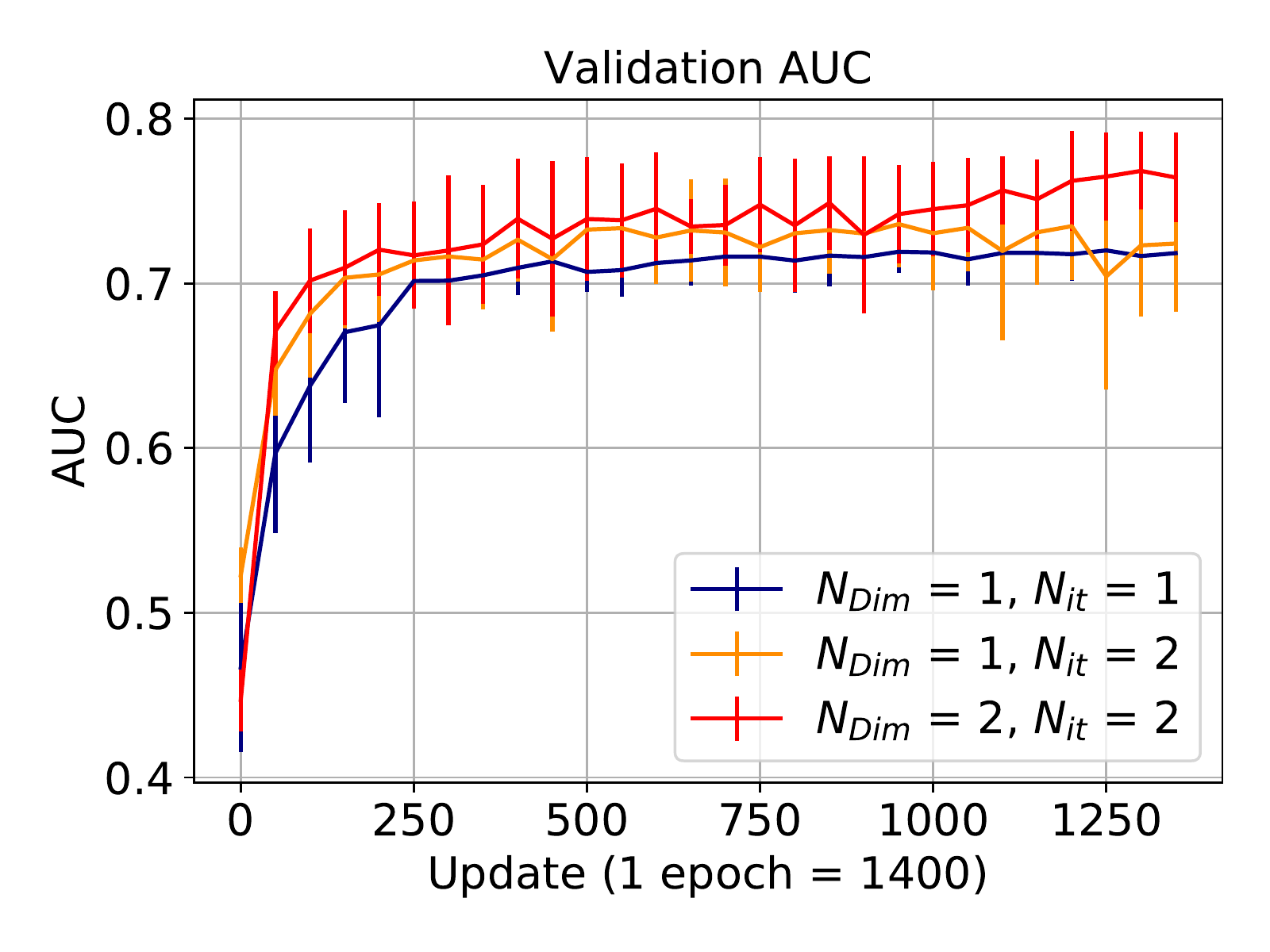}%
\includegraphics[width=0.5\linewidth]{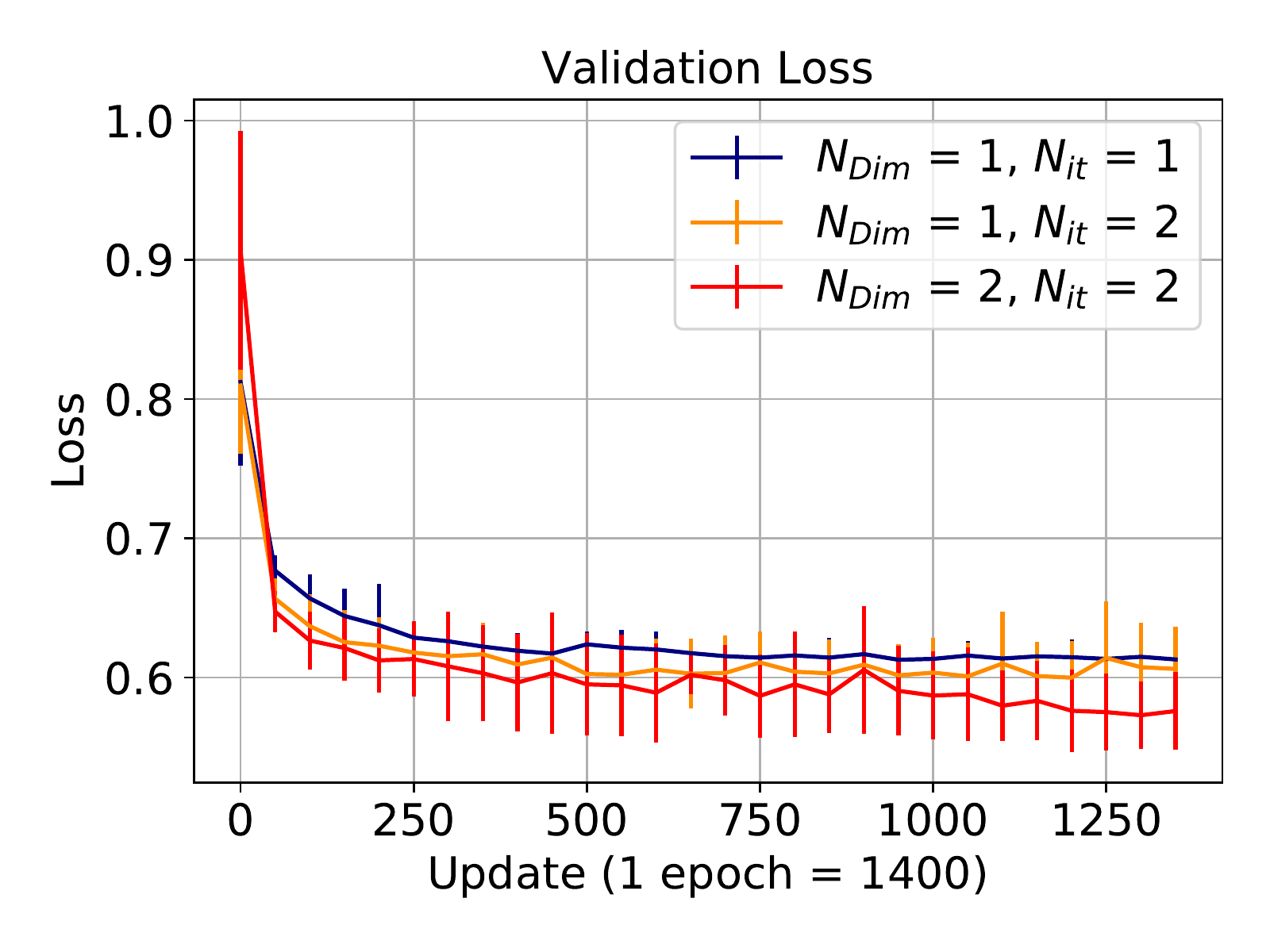}
\caption{Learning curves of the TTN model for different hidden dimensions and iterations. The TTN model performs better as both hidden dimension and iterations increase.  The AUC curve is plotted on the left. The loss curve is plotted on the right.}
\label{fig:results-ttn}
\end{figure}

Comparison against the classical model shows that the QGNN approach is not exceeding the AUC values of the classical model but has a similar performance. There is no clear Quantum Advantage that we can declare with the current results. A better assessment of a possible advantage should be made after exploring the possible benefits of increasing the hidden dimension size and number of iterations. As we have previously shown, increasing the size of the hidden dimension and number of iterations improves the performance significantly~\cite{tuysuz-2}. Results of training a model with different parameters clearly show potential of the improvements provided with increase in hidden dimension size in Fig.~\ref{fig:results-ttn}.

\section{Conclusion}\label{chapter:conclusion}
In this work, we implement a, first of its kind, Quantum Graph Neural Network model for particle track reconstruction. We showed that the simplified model can perform similar to classical approaches. We also gave an evidence on how scaling the network in hidden dimension size can improve its performance. However, it is known that there are many obstacle against training extended Quantum Neural Networks efficiently. Therefore, we keep a conservative attitude on the possible advantages of the Quantum Neural Networks. 

The model is only tested for 1 epoch, which is very small for Deep Learning models. The long simulation time currently prevents us to test more than 1 epoch, as the simulation time takes around a week to complete. Therefore, we wish to improve on this aspect in future work.

The model in this work can be improved on many aspects. Better information encoding strategies that makes better use of the exponentially growing Hilbert State should be followed~\cite{quantum-embedding,quantum-embedding-2}. Also, an extended analysis of PQCs can be made before training~\cite{qc-assesment}. Therefore, PQCs should be selected accordingly. Additionally, strategies to avoid barren plateaus should be followed when training hybrid networks~\cite{barren-2,layerwise,qml-training,qml-training-2}. All of these suggestions apply to almost any hybrid model that uses PQCs. 

This work presents a first complete look on particle track reconstruction using a QGNN. In future work, we plan to improve the current performance considering the items listed above.

\section*{Acknowledgment}

Part of this work was conducted at "\textit{iBanks}", the AI GPU cluster at Caltech. We acknowledge NVIDIA, SuperMicro and the Kavli Foundation for their support of "\textit{iBanks}". This work was partially supported by Turkish Atomic Energy Authority (TAEK) (Grant No: 2017TAEKCERN-A5.H6.F2.15)

\bibliographystyle{IEEEtran}
\bibliography{main}

\begin{thebibliography}{10}
\providecommand{\url}[1]{#1}
\csname url@samestyle\endcsname
\providecommand{\newblock}{\relax}
\providecommand{\bibinfo}[2]{#2}
\providecommand{\BIBentrySTDinterwordspacing}{\spaceskip=0pt\relax}
\providecommand{\BIBentryALTinterwordstretchfactor}{4}
\providecommand{\BIBentryALTinterwordspacing}{\spaceskip=\fontdimen2\font plus
\BIBentryALTinterwordstretchfactor\fontdimen3\font minus
  \fontdimen4\font\relax}
\providecommand{\BIBforeignlanguage}[2]{{%
\expandafter\ifx\csname l@#1\endcsname\relax
\typeout{** WARNING: IEEEtran.bst: No hyphenation pattern has been}%
\typeout{** loaded for the language `#1'. Using the pattern for}%
\typeout{** the default language instead.}%
\else
\language=\csname l@#1\endcsname
\fi
#2}}
\providecommand{\BIBdecl}{\relax}
\BIBdecl

\bibitem{atlas-report}
\BIBentryALTinterwordspacing
``{ATLAS Phase-II Upgrade Scoping Document},'' CERN, Geneva, Switzerland,
  Technical Report CERN-LHCC-2015-020.LHCC-G-166, September 2015. [Online].
  Available: \url{https://cds.cern.ch/record/2055248}
\BIBentrySTDinterwordspacing

\bibitem{cms-report}
``{Technical Proposal for the Phase-II Upgrade of the CMS Detector},'' 6 2015.

\bibitem{lhc-computing}
\BIBentryALTinterwordspacing
J.~Albrecht, A.~A. Alves, G.~Amadio, G.~Andronico, N.~Anh-Ky, L.~Aphecetche,
  J.~Apostolakis, M.~Asai, L.~Atzori, and et~al., ``{A Roadmap for HEP Software
  and Computing R\&D for the 2020s},'' \emph{Computing and Software for Big
  Science}, vol.~3, no.~1, Mar 2019. [Online]. Available:
  \url{http://dx.doi.org/10.1007/s41781-018-0018-8}
\BIBentrySTDinterwordspacing

\bibitem{computing-report}
\BIBentryALTinterwordspacing
D.~Lucchesi, ``{Computing Resources Scrutiny Group Report},'' CERN, Geneva,
  Switzerland, Tech. Rep. CERN-RRB-2017-125, 2017. [Online]. Available:
  \url{http://cds.cern.ch/record/2284575}
\BIBentrySTDinterwordspacing

\bibitem{ref-hilumi}
\BIBentryALTinterwordspacing
G.~Apollinari, O.~Br\"uning, T.~Nakamoto, and L.~Rossi, ``{High Luminosity
  Large Hadron Collider HL-LHC},'' \emph{CERN Yellow Rep.}, no.~5, pp. 1--19,
  2015. [Online]. Available: \url{https://arxiv.org/abs/1705.08830}
\BIBentrySTDinterwordspacing

\bibitem{qml-hep}
\BIBentryALTinterwordspacing
W.~Guan, G.~Perdue, A.~Pesah, M.~Schuld, K.~Terashi, S.~Vallecorsa, and J.-R.
  Vlimant, ``{Quantum Machine Learning in High Energy Physics},'' may 2020.
  [Online]. Available: \url{http://arxiv.org/abs/2005.08582}
\BIBentrySTDinterwordspacing

\bibitem{qalg1}
\BIBentryALTinterwordspacing
I.~Shapoval and P.~Calafiura, ``{Quantum Associative Memory in HEP Track
  Pattern Recognition},'' 2019, preprint:
  \href{https://arxiv.org/abs/1902.00498}{arXiv:quant-ph/1902.00498}. [Online].
  Available: \url{http://arxiv.org/abs/1902.00498}
\BIBentrySTDinterwordspacing

\bibitem{qalg2}
\BIBentryALTinterwordspacing
F.~Bapst, W.~Bhimji, P.~Calafiura, H.~Gray, W.~Lavrijsen, and L.~Linder, ``{A
  pattern recognition algorithm for quantum annealers},'' 2019. [Online].
  Available: \url{http://arxiv.org/abs/1902.08324}
\BIBentrySTDinterwordspacing

\bibitem{qalg3}
\BIBentryALTinterwordspacing
A.~Zlokapa, A.~Anand, J.-R. Vlimant, J.~M. Duarte, J.~Job, D.~Lidar, and
  M.~Spiropulu, ``{Charged particle tracking with quantum annealing-inspired
  optimization},'' 2019. [Online]. Available:
  \url{http://arxiv.org/abs/1908.04475}
\BIBentrySTDinterwordspacing

\bibitem{tuysuz-1}
\BIBentryALTinterwordspacing
C.~Tüysüz, F.~Carminati, B.~Demirköz, D.~Dobos, F.~Fracas, K.~Novotny,
  K.~Potamianos, S.~Vallecorsa, and J.-R. Vlimant, ``{Particle Track
  Reconstruction with Quantum Algorithms},'' 2020. [Online]. Available:
  \url{https://arxiv.org/abs/2003.08126}
\BIBentrySTDinterwordspacing

\bibitem{tuysuz-2}
\BIBentryALTinterwordspacing
C.~Tüysüz \emph{et~al.}, ``{A Quantum Graph Neural Network Approach to
  Particle Track Reconstruction},'' 2020. [Online]. Available:
  \url{https://arxiv.org/abs/2007.06868}
\BIBentrySTDinterwordspacing

\bibitem{trackml}
S.~Amrouche, L.~Basara, P.~Calafiura, V.~Estrade, S.~Farrell, D.~R. Ferreira,
  L.~Finnie, N.~Finnie, C.~Germain, V.~V. Gligorov, T.~Golling, S.~Gorbunov,
  H.~Gray, I.~Guyon, M.~Hushchyn, V.~Innocente, M.~Kiehn, E.~Moyse, J.-F.
  Puget, Y.~Reina, D.~Rousseau, A.~Salzburger, A.~Ustyuzhanin, J.-R. Vlimant,
  J.~S. Wind, T.~Xylouris, and Y.~Yilmaz, ``The tracking machine learning
  challenge: Accuracy phase,'' in \emph{The {NeurIPS} 2018 Competition}.\hskip
  1em plus 0.5em minus 0.4em\relax Springer International Publishing, Nov.
  2019, pp. 231--264.

\bibitem{HepTrkX}
\BIBentryALTinterwordspacing
S.~Farrell \emph{et~al.}, ``{Novel deep learning methods for track
  reconstruction},'' 2018. [Online]. Available:
  \url{http://arxiv.org/abs/1810.06111}
\BIBentrySTDinterwordspacing

\bibitem{barren}
\BIBentryALTinterwordspacing
J.~R. McClean, S.~Boixo, V.~N. Smelyanskiy, R.~Babbush, and H.~Neven, ``{Barren
  plateaus in quantum neural network training landscapes},'' \emph{Nat.
  Commun.}, vol.~9, no.~1, pp. 1--6, 2018. [Online]. Available:
  \url{http://dx.doi.org/10.1038/s41467-018-07090-4}
\BIBentrySTDinterwordspacing

\bibitem{barren-2}
\BIBentryALTinterwordspacing
E.~Grant, L.~Wossnig, M.~Ostaszewski, and M.~Benedetti, ``{An initialization
  strategy for addressing barren plateaus in parametrized quantum circuits},''
  \emph{Quantum}, vol.~3, p. 214, 2019. [Online]. Available:
  \url{https://arxiv.org/abs/1903.05076}
\BIBentrySTDinterwordspacing

\bibitem{qpic}
\BIBentryALTinterwordspacing
``$\bra{q}\ket{pic}$.'' [Online]. Available: \url{https://github.com/qpic/qpic}
\BIBentrySTDinterwordspacing

\bibitem{mps}
\BIBentryALTinterwordspacing
A.~S. Bhatia, M.~K. Saggi, A.~Kumar, and S.~Jain, ``{Matrix product
  state–based quantum classifier},'' \emph{Neural Computation}, vol.~31,
  no.~7, pp. 1499--1517, 2019. [Online]. Available:
  \url{https://arxiv.org/abs/1905.01426}
\BIBentrySTDinterwordspacing

\bibitem{ttn}
E.~Grant, M.~Benedetti, S.~Cao, A.~Hallam, J.~Lockhart, V.~Stojevic, A.~G.
  Green, and S.~Severini, ``{Hierarchical quantum classifiers},'' \emph{npj
  Quantum Information}, vol.~4, no.~1, pp. 17--19, 2018.

\bibitem{parameter-shift}
\BIBentryALTinterwordspacing
M.~Schuld, V.~Bergholm, C.~Gogolin, J.~Izaac, and N.~Killoran, ``{Evaluating
  analytic gradients on quantum hardware},'' \emph{Phys. Rev. A}, vol.~99, p.
  032331, Mar 2019. [Online]. Available:
  \url{https://link.aps.org/doi/10.1103/PhysRevA.99.032331}
\BIBentrySTDinterwordspacing

\bibitem{ref-tensorflow}
\BIBentryALTinterwordspacing
M.~Abadi \emph{et~al.}, ``{TensorFlow}: Large-scale machine learning on
  heterogeneous systems,'' 2015, software available from tensorflow.org.
  [Online]. Available: \url{https://www.tensorflow.org/}
\BIBentrySTDinterwordspacing

\bibitem{pennylane}
\BIBentryALTinterwordspacing
V.~Bergholm, J.~Izaac, M.~Schuld, C.~Gogolin, C.~Blank, K.~McKiernan, and
  N.~Killoran, ``{PennyLane: Automatic differentiation of hybrid
  quantum-classical computations},'' pp. 1--12, 2018. [Online]. Available:
  \url{http://arxiv.org/abs/1811.04968}
\BIBentrySTDinterwordspacing

\bibitem{qulacs}
\BIBentryALTinterwordspacing
``Qulacs.'' [Online]. Available: \url{https://github.com/qulacs/qulacs}
\BIBentrySTDinterwordspacing

\bibitem{quantum-embedding}
\BIBentryALTinterwordspacing
S.~Lloyd, M.~Schuld, A.~Ijaz, J.~Izaac, and N.~Killoran, ``Quantum embeddings
  for machine learning,'' 2020. [Online]. Available:
  \url{https://arxiv.org/abs/2001.03622}
\BIBentrySTDinterwordspacing

\bibitem{quantum-embedding-2}
\BIBentryALTinterwordspacing
M.~Schuld, R.~Sweke, and J.~J. Meyer, ``The effect of data encoding on the
  expressive power of variational quantum machine learning models,'' 2020.
  [Online]. Available: \url{https://arxiv.org/abs/2008.08605}
\BIBentrySTDinterwordspacing

\bibitem{qc-assesment}
\BIBentryALTinterwordspacing
S.~Sim, P.~D. Johnson, and A.~Aspuru‐Guzik, ``"expressibility and entangling
  capability of parameterized quantum circuits for hybrid quantum‐classical
  algorithms",'' \emph{Advanced Quantum Technologies}, vol.~2, no.~12, p.
  1900070, Oct 2019. [Online]. Available:
  \url{http://dx.doi.org/10.1002/qute.201900070}
\BIBentrySTDinterwordspacing

\bibitem{layerwise}
\BIBentryALTinterwordspacing
A.~Skolik, J.~R. McClean, M.~Mohseni, P.~van~der Smagt, and M.~Leib,
  ``{Layerwise learning for quantum neural networks},'' 2020. [Online].
  Available: \url{https://arxiv.org/abs/2006.14904}
\BIBentrySTDinterwordspacing

\bibitem{qml-training}
\BIBentryALTinterwordspacing
A.~Mari, T.~R. Bromley, and N.~Killoran, ``Estimating the gradient and
  higher-order derivatives on quantum hardware,'' 2020. [Online]. Available:
  \url{https://arxiv.org/abs/2008.06517}
\BIBentrySTDinterwordspacing

\bibitem{qml-training-2}
\BIBentryALTinterwordspacing
M.~Cerezo and P.~J. Coles, ``Impact of barren plateaus on the hessian and
  higher order derivatives,'' 2020. [Online]. Available:
  \url{https://arxiv.org/abs/2008.07454}
\BIBentrySTDinterwordspacing

\end{thebibliography}
\end{document}